\documentclass[12pt]{article}
\usepackage{amsmath,amssymb,bbold,epsfig}

\newcommand{\beq}{\begin{equation}}
\newcommand{\be}{\begin{equation}}
\newcommand{\ee}{\end{equation}}
\newcommand{\bea}{\begin{eqnarray}}
\newcommand{\eea}{\end{eqnarray}}

\begin{document}
\topmargin 0pt  
\oddsidemargin=-0.4truecm
\evensidemargin=-0.4truecm

\setcounter{page}{1}
\begin{titlepage}  
\vspace*{-2.0cm}
\begin{flushright}
hep-ph/0610198
\end{flushright}
\vspace*{0.2cm}
\begin{center}
\vspace*{0.1cm}
{\Large \bf Neutrino oscillations: what is magic about the ``magic'' 
baseline?}
\\
\vspace{0.8cm}  
{A. Yu. Smirnov \footnote{E-mail address:
smirnov@ictp.trieste.it}}
\vspace{0.2cm}

{\em The Abdus Salam International Centre for Theoretical
Physics \\
Strada Costiera 11, I-34014 Trieste, Italy}
\vspace*{0.1cm}

{\em  Institute for Nuclear Research, Russian Academy of 
Sciences \\
 Moscow, Russia}

\vspace*{0.1cm}

\end{center}
\vglue 0.3truecm

\begin{abstract}
Physics interpretation of the ``magic'' baseline, 
$L_{magic}$, that can play important role 
in future oscillation experiments  is given. The ``magic'' baseline 
coincides with the refraction length, $l_0$. 
The latter, in turn, approximately equals the oscillation length in matter at   
high energies. Therefore  at the baseline $L = l_0$  
the oscillation phase is $2\pi$, and consequently, 
the ``solar'' amplitude of oscillations driven  
by $\theta_{12}$ and $\Delta m^2_{21}$ vanishes. 
As a result, in the lowest order  (i) the  interference of  
amplitudes in the  $\nu_e - \nu_\mu$ $(\nu_\tau)$ transition probability  
is absent; (ii)  dependence of the probability on the CP-phase, $\delta$, 
as well as  on $\theta_{12}$ and  $\Delta m^2_{21}$ disappears. 
Corrections to the equality $L_{magic} = l_0$  are estimated. 
Effect of changing density is considered and two new magic 
trajectories are identified for neutrinos that cross  the core of the 
Earth.  Other magic baselines associated with zeros of the atmospheric 
amplitude are discussed. 

\end{abstract}

\vspace{1.cm}
\vspace{.3cm}

\end{titlepage}
\renewcommand{\thefootnote}{\arabic{footnote}}
\setcounter{footnote}{0}
\newpage

\section{Introduction}

It was observed some time ago that at  the baseline 
\be
L_{magic} = \frac{2\pi}{\sqrt{2}G_F n_e }, 
\label{magic}
\ee
where $G_F$ is the Fermi coupling constant and 
$n_e$ is the electron number density,  
the analytic formula for the $\nu_\mu - \nu_e$ oscillation probability 
in matter (in $3\nu-$ mixing context) takes a very simple form~\cite{barger}. 
The probability does not depend 
on the CP violation phase, $\delta$, as well as on the 
mixing angle, $\theta_{12}$, and the mass splitting, $\Delta m_{21}^2$,  
of the 1-2 sector. Therefore, neutrino oscillation experiments with 
the baseline $L = L_{magic} = (7300 - 7600)$ km  will allow  one 
to perform clean measurements of  $\theta_{13}$  
resolving degeneracy with the phase $\delta$ \cite{barger,winter}.
That was discussed in a number of recent recent publications 
in some details~\cite{more}. 
The baseline $L_{magic}$ depends only on  matter density and does not depend 
on neutrino energy and oscillation parameters. For this reason 
it was termed the ``magic''  baseline in \cite{huber}. 

Apparently the ``magic'' baseline defined in (\ref{magic}) 
coincides with the refraction length, $l_0$, introduced 
by Wolfenstein almost 30 years ago \cite{w}: 
\be
L_{magic} \equiv l_0. 
\label{equal-l}
\ee
This is not accidental coincidence and in what follows we will explain 
the reason behind the equality (\ref{equal-l}). 
The explanation is given in secs. 2 and 3. Secs. 4 and 5 
contain some more advanced material, in particular, 
discussion of various corrections to the equality (\ref{equal-l}).

\section{Magic baseline and refraction length}

Recall that the refraction length, $l_0$, has been defined as the distance 
over which an additional phase difference, $\phi_m$, acquired by neutrinos 
due to interactions with matter,  equals $2\pi$: 
\be
\phi_m = V l_0 = 2\pi. 
\ee
Here  $V \equiv \sqrt{2} G_F n_e$ is the difference of potentials 
for $\nu_e$ and $\nu_{\mu}$ in usual matter, so that  
$l_0$ is a characteristic of medium 
relevant for $\nu_e - \nu_{\mu}$  mixing. 

The refraction length enters the MSW-resonance condition \cite{MSW}, 
\be
l_0 \cos 2\theta = l_{\nu}, 
\label{res}
\ee
where $l_\nu  \equiv  4\pi E/\Delta m^2$ is the vacuum oscillation length and 
$\theta$ is the vacuum mixing angle. 
Inverse quantity, $1/l_0$, determines the eigenfrequency of medium, and   
for small mixing the condition (\ref{res}) means  that 
the eigenfrequency of medium  coincides with 
the eigenfrequency of neutrino system, $1/l_{\nu}$. 
For large mixing (strongly coupled system) there is a shift of 
resonance frequency given by $\cos 2\theta$. 

The ratio  $l_0/l_{\nu}$ determines modifications of the  
mixing angle, $\theta_m$, and oscillation length in matter $l_m$: 
\be 
\sin 2\theta_m =  \sin 2\theta [(\cos 2 \theta  - l_\nu/l_0)^2 + \sin^2 2 \theta]^{-1/2}.
\label{angle}
\ee
\be 
l_m = l_\nu [(\cos 2 \theta  - l_\nu/l_0)^2 + \sin^2 2 \theta]^{-1/2}.
\label{length}
\ee
According to (\ref{length}),  at low energies (below resonance)
one has $l_m \approx l_\nu$; with increase of energy the length 
$l_m$ first increases, it  reaches 
maximal value, $l_m^{max} = l_0/\sin 2\theta$, at $l_{\nu} = l_0/\cos 2\theta$ 
(i.e., above the resonance: $E^{max} = E_R/\cos^2 2\theta$) 
\footnote{I am grateful to J. Kersten for correcting the maximal value.},  
and then  decreases approaching $l_0$ from above. 
In the non-resonance channel,  $l_m$ 
increases with $E$  approaching  $l_0$ from below.  
Thus, in the case of large densities of matter or large energies of neutrinos,  
when $l_\nu \gg l_0$, we obtain 
\be 
l_m \approx l_0. 
\ee
In other words, in the matter dominating case 
the oscillation length in matter approximately equals
the refraction length. This is the key point of physics interpretation of the 
magic baseline in the next section.      

\section{Where does the magic baseline come from?
\label{why}
}

Let us consider oscillations of three mixed neutrinos  
$\nu_f \equiv (\nu_e, \nu_{\mu}, \nu_{\tau})^T$
in the matter.  The vacuum mixing matrix, $U_{PMNS}$, that  relates $\nu_f$ and 
the mass 
eigenstates $\nu = (\nu_1, \nu_2, \nu_3)^T$, $\nu_f = U_{PMNS} \nu$, 
can be parametrized as    
\be
U_{PMNS} = U_{23} I_{\delta} U_{13} I_{-\delta}U_{12}.    
\label{mixing}
\ee
Here $U_{ij}= U_{ij}(\theta_{ij})$ performs  rotation  
in the $ij$- plane by the angle $\theta_{ij}$ 
and $I_{\delta} \equiv diag(1, 1, e^{i\delta})$ 
is the matrix of CP-violation phase. 

The  $\nu_\mu \to \nu_e$  transition probability  
can be represented as 
\be
P(\nu_\mu \to \nu_e) = |\cos \theta_{23} A_{S} e^{i\delta} + 
\sin \theta_{23} A_{A}|^2
\label{prob-mue}
\ee
(see some details in sec. 4). 
Here $A_{S}$ is the solar amplitude that in the lowest order approximation 
(up to corrections of the order 
$\Delta m_{21}^2/\Delta m_{31}^2$, and $\sin^2 \theta_{13}$) 
depends on the solar neutrino oscillation parameters: 
$\Delta m_{21}^2$, $\theta_{12}$.  
$A_{A}$ is the atmospheric amplitude that  depends on the 
atmospheric neutrino oscillation parameters,  
$\Delta m_{31}^2$, $\theta_{13}$. 
The CP-violation effects are given by interference of the two amplitudes 
in (\ref{prob-mue}). 

Let us consider the constant  density medium  that, in fact, is a very 
good approximation for neutrinos propagating in the mantle of the Earth.  
Up to the phase factor we obtain for the uniform medium 
\be 
A_S = \sin 2 \theta_{12}^m \sin \frac{\phi_S^m}{2},   
\label{ampl-S}
\ee
where $\theta_{12}^m$ is the 1-2 mixing angle (\ref{angle}) 
and $\phi_S^m$ is the oscillation phase in matter: 
\be
\phi_S^m \equiv  \frac{2\pi L}{l_m};
\ee
the oscillation length, $l_m$, is given in (\ref{length}). 
Apparently $A_S$ is a square root of the usual oscillation probability. 
(Similar expression can be written for $A_A$ with substitution 
1-2 parameters by 1-3 oscillation parameters.)

Let us consider neutrinos with energies $E > (0.5 - 1)$ GeV 
relevant for accelerator  experiments and atmospheric neutrino studies. 
For these energies 
$l_{\nu} (\Delta m_{21}^2)\equiv 4\pi E/\Delta m_{21}^2 \gg l_0$, 
and therefore the solar oscillation mode 
is in the matter dominating regime when 
\be
l_m (\Delta m_{21}^2) \approx l_0.
\ee
Correspondingly, the phase of oscillations equals 
\be
\phi_S^m \approx \frac{2\pi L}{l_0},
\ee
and  at the baseline  $L = l_0$  we obtain
\be
\phi_S^m = 2\pi, ~~~  A_{S} = 0. 
\ee
The solar amplitude vanishes,  and consequently, 
\be
P(\nu_\mu \to \nu_e) \approx |\sin \theta_{23} A_{A}|^2. 
\ee
So, for high energies and $L = l_0$ the ``magic'' properties are reproduced:  

- the interference and dependence on the phase $\delta$ disappear; 

- the dependence on solar oscillation parameters disappear too. 

Summarizing, {\it the ``magic'' length is nothing but the refraction length}, 
at least in the lowest order in  small parameters.  
For high energies  at the distance $L = l_0$,  
the phase of ``solar'' oscillation amplitude  
becomes $2\pi$. The amplitude 
vanishes,  and therefore  dependence of probability on CP-phase disappears.
Similar consideration is valid for the $\nu_e - \nu_{\tau}$ channel with 
substitution $\sin \theta_{23} \rightarrow \cos \theta_{23}$ and 
$\cos \theta_{23} \rightarrow - \sin  \theta_{23}$ in 
(\ref{prob-mue}). 

\section{Corrections to the equality $L_{magic} = l_0$
\label{corrections
}}

In the previous section we have neglected terms of the 
order $\sin^2 \theta_{13}$ and $\Delta m^2_{21}/\Delta m^2_{31}$  
which can be as large as  $3\%$. Also we have taken 
$\l_m = l_0$, but at relatively low energies the deviation 
from this equality can be significant.   
Notice that the $3\%$ correction to the refraction length of 7300 km 
equals 220 km which may be non-negligible for selection of a detector place, 
if high precision measurements of parameters are planned. 
One way to proceed is to perform numerical search of the baseline  when 
the probability has the weakest dependence on $\delta$ \cite{winter}. 
Here we present some analytic consideration. \\

We will define the magic baseline as the distance over which 
the ``solar'' amplitude vanishes:  
\be
A_S(L_{magic}) = 0. 
\ee
In the constant density approximation (and in the lowest order in $\sin \theta_{13}$) 
this leads to the condition $L_{magic} = l_m$, and for high energies $l_m \approx l_0$. 
There are two types of corrections to the equality $L_{magic} \approx l_0$: 
(i) due deviation of $l_m$ from  the asymptotic value,  $l_0$, and (ii) due to 
dependence of $A_S$ on $\sin \theta_{13}$. We consider these corrections 
in order.

1). Corrections due to  $l_m \neq l_0$: For $l_\nu \gg l_0$ we have 
\be
L_{magic} = l_m \approx l_0 \left(1  + \cos 2 \theta_{12} \frac{l_0}{l_\nu} \right) = 
l_0 \left(1  + \cos 2 \theta_{12} \frac{\Delta m_{21}^2}{2EV} \right) 
\label{corr1}
\ee
in  the resonance channel. 
The correction is positive and equals $\approx 0.1/E$ (GeV). 
For a constant density medium,   
with increase of energy the correction and the magic baseline become smaller.  
However, in the case of realistic Earth density profile an average 
density along a trajectory increases with a length of trajectory, or equivalently, with 
$|\cos \Theta_\nu|$, where $\Theta_\nu$ is the zenith  angle of the trajectory.  
Therefore $l_0$ decreases. So, for the Earth 
the magic trajectory does not change significantly with the energy being 
at $|\cos \Theta_\nu| \approx 0.6$.  
For the non-resonance channel the correction is negative (minus sign in 
Eq. (\ref{corr1})). \\

2). Corrections due to 1-3 mixing. 
To evaluate these corrections 
we need to give precise definitions   
of the amplitudes $A_S$ and $A_A$ in (\ref{prob-mue}) 
and find their dependence on the oscillation parameters. The evolution of the flavor states 
$\nu_f$ is described by the Hamiltonian 
\be
H = U_{PMNS} Diag(0, \Delta_{21}, \Delta_{31}) U_{PMNS}^{\dagger} + \hat{V}, 
\label{flH}
\ee
where $\hat{V} = diag(V, 0, 0)$,  $\Delta_{21} \equiv \Delta m_{21}^2/2E$ and  $\Delta_{31} 
\equiv \Delta m_{31}^2/2E$. 
Let us define the neutrino propagation basis,  
$\tilde{\nu} = (\nu_e, \tilde{\nu}_{2}, \tilde{\nu}_{3})^T$,   
through   
\be
\nu_f = U_{23}I_{\delta} \tilde{\nu}. 
\label{basisrel}
\ee
The Hamiltonian $\tilde{H}$ that 
describes oscillations of $\tilde{\nu}$  can be obtained 
from  (\ref{flH}), (\ref{mixing}) and (\ref{basisrel}): 
$$
\tilde{H} =  U_{13} U_{12} Diag(0, \Delta_{21}, \Delta_{31})
U^\dagger_{12} U^\dagger_{13} ~ + ~ \hat{V},
$$ 
or explicitly, 
\be
\tilde{H} =   
\left(\begin{array}{ccc}
c_{13}^2 s_{12}^2 \Delta_{12} + V + s_{13}^2 \Delta_{13} & c_{13} s_{12} c_{12} \Delta_{12}
& s_{13}c_{13} \Delta_{13} - s_{13}c_{13} s_{12}^2 \Delta_{12}  \\
... & c_{12}^2 \Delta_{12}   &   - s_{13} s_{12} c_{12} \Delta_{12}   \\
... &  ...  &  c_{13}^2 \Delta_{13} + s_{13}^2 s_{12}^2 \Delta_{12}
\end{array}\right)\, ,   
\label{matr1}
\ee
where $c_{13} \equiv \cos \theta_{13}$, $s_{13} \equiv \sin \theta_{13}$, 
{\it etc.}.   
Introducing the evolution matrix in the propagation basis as 
\be
\tilde{S} = || A_{ij}|| , ~~~ i, j = e, 2, 3, 
\ee
it is straightforward to find that the $\nu_\mu - \nu_e$ transition  
probability has the form (\ref{prob-mue}) with 
\be
A_{S} = A_{e2}, ~~~ A_{A} =  A_{e3}. 
\label{connect}
\ee
That is,  the solar amplitude is given by the amplitude of transition  
$\tilde{\nu}_{2} \rightarrow \nu_e$ 
and the atmospheric amplitude coincides with 
the  $\tilde{\nu}_{3} \rightarrow \nu_e$ transition amplitude.  In the 
limit $\theta_{13} \rightarrow 0$ the state  
$\tilde{\nu}_{3}$ becomes the mass eigenstate $\nu_3$
and decouples from the rest of neutrino system,  
whereas $\tilde{\nu}_{2}$  becomes the combination 
of $\nu_{\mu}$ and $\nu_{\tau}$ that  mixes with $\nu_e$ 
with the solar oscillation parameters. 
Therefore in this limit $A_{e2} = A_{e2} (\Delta m_{21}^2, \theta_{12})$
as we discussed in sec. \ref{why}. 

For energies  below the 1-3 resonance ($E \sim 1$ GeV), when the 
33-element of the Hamiltonian dominates, the corrections  can  
be calculated immediately. Performing an additional rotation of the neutrino basis 
by $U_{13}(\theta_{13})$ and then making block-diagonalization of the obtained 
Hamiltonian,  
we obtain that the heaviest state decouples and the two others form  
a $2\nu-$ system with the mixing angle $\theta_{12}$ and modified potential: 
\be
V \rightarrow V - s_{13}^2 V\left(1 + \frac{V}{\Delta_{13}} \right).
\ee
Correspondingly, the magic length is modified as 
\be
L_{magic} \approx l_0 \left[ 1 + \cos 2\theta_{12} \frac{l_0}{l_{\nu}} + 
s_{13}^2  \right]. 
\label{magcor2}
\ee
Notice that the block-diagonalization removes the  imaginary 
part of the corrections to the solar amplitude (see discussion below).\\ 

Let us present an estimation of the $s_{13}$-corrections that are valid 
in whole the energy range including the 1-3 resonance, and also 
take into account the imaginary part of the solar amplitude. 
We  perform an additional 1-3 rotation of the neutrino basis 
that vanishes the 1-3 element of the Hamiltonian (\ref{matr1}): 
\be
\tilde \nu = U_{13} (\theta_{13}^m) \nu_m. 
\ee
Here $\nu_m \equiv (\nu_{1m}, \tilde{\nu}_2, \nu_{3m})$, and the angle is given by 
\be
\tan 2\theta_{13}^m = \frac{2\tilde{H}_{e3}}{\tilde{H}_{33} - \tilde{H}_{ee}},  
\label{mix13mat}
\ee
with  $\tilde{H}_{ij}$ being the $ij$-element of the Hamiltonian (\ref{matr1}). 
In the new basis, the Hamiltonian has the form 
\be
{H}^m = 
\left(\begin{array}{ccc}
H_{11}^m &  \cos (\theta_{13}^m - \theta_{13}) a_{12}  & 0  \\
... & \tilde{H}_{22}   &   \sin (\theta_{13}^m - \theta_{13}) a_{12} \\
... &  ...  &  H_{33}^m
\end{array}\right)\, ,
\label{matr2}
\ee
where 
\be
a_{12} \equiv s_{12} c_{12} \Delta_{12}, 
\ee 
\begin{eqnarray}
H_{11}^m = \cos^2 \theta_{13}^m \tilde{H}_{ee} - \sin 2\theta_{13}^m \tilde{H}_{e3}
+ \sin^2 \theta_{13}^m \tilde{H}_{33}, 
\nonumber\\
H_{33}^m = \sin^2 \theta_{13}^m \tilde{H}_{ee} + \sin 2\theta_{13}^m \tilde{H}_{e3}
+ \cos^2 \theta_{13}^m \tilde{H}_{33}. 
\end{eqnarray}
Since we are looking for the corrections due to the 1-3 mixing the terms 
$\propto \Delta_{12}$ can be omitted in the diagonal elements, 
$H_{11}^m$,  $H_{33}^m$ , and  consequently we obtain 
\be 
H_{11}^m = 
V \frac{\cos \theta_{13} \cos \theta_{13}^m}{\cos(\theta_{13}^m -\theta_{13})}, 
~~~H_{33}^m = 
V \frac{\cos \theta_{13} \sin \theta_{13}^m}{\sin(\theta_{13}^m -\theta_{13})}. 
\label{diag-el}
\ee 
Here we used expression for $\Delta_{13}$ in terms of mixing angle in matter 
that can be obtained from (\ref{mix13mat}): 
\be 
\Delta_{13} = \frac{V \sin 2\theta_{13}^m}{\sin2(\theta_{13}^m -\theta_{13})}. 
\ee

In the propagation basis, the transition  $\nu_e \rightarrow  \tilde{\nu}_{2}$  proceeds  
in two different ways: 
$\nu_e \rightarrow \nu_{1m} \rightarrow  \tilde{\nu}_{2}$ and $\nu_e \rightarrow  
\nu_{3m} \rightarrow  \tilde{\nu}_{2}$. 
Therefore  the ``solar'' amplitude can be written as  
\be
A_{e2} = \cos \theta_{13}^m A_{12}^m + \sin \theta_{13}^m A_{32}^m, 
\label{amplcor}
\ee
where $A_{12}^m$ and $A_{32}^m$ 
are the amplitudes of $\nu_{1m} \rightarrow  \tilde{\nu}_{2}$ 
and  $\nu_{3m} \rightarrow  \tilde{\nu}_{2}$ 
transitions correspondingly. Then  
in the lowest approximation using the Hamiltonian (\ref{matr2}) we obtain for 
(\ref{amplcor}): 
\begin{eqnarray}
A_{e2} \approx  2 a_{12} \left[
\cos \theta_{13}^m  \cos(\theta_{13}^m -\theta_{13})  
\frac{\sin (H^m_{11}L/2)}{H^m_{11}} e^{-i H^m_{11} L/2}  \right.
\nonumber \\
+ \left. \sin \theta_{13}^m \sin (\theta_{13}^m -\theta_{13})   
\frac{\sin (H^m_{33}L/2)}{H^m_{33}} e^{-i H^m_{33}L/2}
\right].
\label{amplcor3}
\end{eqnarray}
For energies much below the 1-3 resonance  we have $\sin \theta_{13}^m \approx 
\sin \theta_{13} \ll 1$. Therefore the second term 
in (\ref{amplcor3}) is strongly suppressed, furthermore $H^m_{11} \approx \tilde{H}_{ee}
\approx V$, and consequently, the amplitude is reduced to the one considered in sec. 3 is 
recovered. For energies above the 1-3 resonance, $\cos \theta_{13}^m \rightarrow 0$ 
and the first term
in (\ref{amplcor}) vanishes. Now $H^m_{33} \approx \tilde{H}_{ee}
\approx V$ and again we recover the result of sec. 3.

Minimal value of $A_{e2}$ corresponds to 
$H_{11}^m L \approx H_{33}^m L \approx 2\pi$ when both terms in 
(\ref{amplcor3}) are close to zero. 
(We confirm this by explicit 
calculation.) 
Then introducing small quantities 
\be 
\epsilon_1 = \frac{1}{2} H_{11}^m L - \pi, ~~~~
\epsilon_3 = \frac{1}{2}  H_{33}^m L  - \pi   
\ee
($\epsilon_i \ll \pi$), 
and taking the first 
terms of expansions of sines and exponents in (\ref{amplcor3}) 
around $\pi$ we obtain: 
\begin{eqnarray}
A_{e2} \approx  2 a_{12} \left[
\cos \theta_{13}^m  \cos(\theta_{13}^m -\theta_{13})
(1 - i \epsilon_1) \left(\frac{L}{2} - \frac{\pi}{H_{11}^m}\right) \right.
\nonumber \\
+ \left. \sin \theta_{13}^m \sin (\theta_{13}^m -\theta_{13})
(1 - i \epsilon_3) \left(\frac{L}{2} - \frac{\pi}{H_{33}^m}\right)
\right].
\label{amplcor5}
\end{eqnarray}
The imaginary part contains an additional power of small parameters 
$\epsilon_i$. 

Using explicit expressions for the $H_{11}^m$
and $H_{33}^m$ (\ref{diag-el}) we obtain from (\ref{amplcor5}) the real and imaginary 
parts of the amplitude: 
\be
A_{e2}^{(R)}  = c_{13} s_{12} c_{12} \Delta_{12}  
\left(L - \frac{2\pi}{V c_{13}^2}\right), 
\label{realamp}
\ee
\be
A_{e2}^{(I)}  =  - \frac{1}{2} c_{13} s_{12} c_{12} \Delta_{12} V
\left[ \left(L - \frac{2\pi}{V}\right)^2 + \frac{4\pi^2}{V^2}\tan^2\theta_{13}\right]. 
\label{imamp}
\ee
Let us analyze these results. 

1). The real part of the amplitude vanishes if 
\be 
L = \frac{2\pi}{V c_{13}^2}
\label{magreal}
\ee
that coincides with the baseline  obtained in (\ref{magcor2}). 

2). The imaginary part is always non-zero with minimum 
\be
|A_{e2}^{(I)}|_{min}  = 2\pi^2  s_{13}^2 s_{12} c_{12} 
\frac{\Delta_{12}}{V}
\label{imampmin}
\ee
at $L = {2\pi}/{V}$. 

3). At the baseline that corresponds to  zero real part (\ref{magreal}),  we obtain 
from (\ref{imamp}) that the correction to the minimal value 
(\ref{imampmin}) is of the order $s_{13}^4$: 
$ |A_{e2}^{(I)}| =  |A_{e2}^{(I)}|_{min} (1 + s_{13}^2)$. 
This means that  the baseline (\ref{magreal}) provides a minimum of the total 
amplitude in the order $s_{13}^2$, 
and therefore it can be identified with the magic length. 
When corrections due to 1-3 mixing are included, 
the solar amplitude does not vanish exactly, and therefore the magic properties 
are satisfied only approximately. 

4). The amplitudes (\ref{realamp}, \ref{imamp}) do not depend on the mixing angle in 
matter, and   therefore the results are valid in whole energy range including the 1-3 resonance 
region and the region above the resonance.

\section{The case of non-constant density}

In the case of non-constant potential (density) along the neutrino 
trajectory, the refraction length can be defined  by the condition  
\be
\int_0^{l_0} dx V(x) = 2\pi.
\ee
In the lowest order, 
for the ``solar'' amplitude in the matter dominating case one can 
obtain the following expression (see \cite{ams} for details)  
\be
A_S = \frac{1}{2} \sin 2\theta_{12} \Delta_{12} \int_0^L dx 
 \exp\left(-i \int_x^L dy V(y) \right).   
\label{sol-gen}
\ee
Consequently,  the magic baseline can be found from the condition
\be
\int_0^{L_{magic}} dx \exp\left(i \int_0^{x} dy V(y) \right) = 0.  
\label{mag-gen}
\ee
The double integration takes into account the change of 
both oscillation length and  mixing angle  along the trajectory. 

Let us consider the three-layer profile with  constant potentials 
(densities) $V_m$, $V_c$ and  $V_m$, and baselines $L_m$, $L_c$ and $L_m$.  
The densities and baselines  in the first and the third layers 
coincide.  To a good approximation 
that corresponds to profile along the neutrino trajectories that cross  
the core of the earth with $V_m$ and  $V_c$ being the potentials in the mantle 
and the core correspondingly. 

Performing integration in eq.  (\ref{sol-gen}) we find
\be 
A_S = \sin 2 \theta_m^0 \left[ \sin \left( \frac{\phi_c}{2} + \phi_m 
\right)
- \left(1 - \frac{V_m}{V_c} \right) \sin\frac{\phi_c}{2} \right], 
\label{solar-gen2}
\ee
where 
\be 
\sin 2 \theta_m^m \approx \sin 2 \theta_{12} \frac{\Delta_{12}}{ V_m} 
\ee
is the mixing angle in 
the mantle, and 
\be
\phi_m = V_m L_m, ~~~~\phi_c = V_c L_c 
\label{phases2}
\ee
are the phases acquired in the mantle and the core.   
The sum   
$\phi_c + 2\phi_m$ is the total oscillation phase. 
The  first term in (\ref{solar-gen2}) corresponds to the amplitude of  
pure adiabatic transition.  
It depends on the mixing angle at the  surface 
of the earth  and on the total phase. 
The second term is the correction due to the adiabaticity violation  at the 
border  between the mantle and the core. 
So, in the presence of the adiabaticity violation the 
amplitude is not  determined by total phase.  

According to (\ref{solar-gen2}) the magic baseline is determined by the 
condition 
\be
\sin \left( \frac{\phi_c}{2} + \phi_m \right) = 
\left(1 - \frac{V_m}{V_c}\right) \sin \frac{\phi_c}{2}. 
\label{magic-gen3}
\ee
For constant density along whole trajectory 
($V_c = V_m$) or in the adiabatic case,  
eq. (\ref{magic-gen3}) would lead to 
\be
\frac{\phi_c}{2} + \phi_m  = \pi k, ~~~~ k = 1, 2, 3...~. 
\label{cond-ad}
\ee
The violation of adiabaticity modifies this condition, 
and apparently, the bigger the difference of 
the potentials $V_m$ and $V_c$ the stronger the deviation from (\ref{cond-ad}). 

In the case of the Earth profile,  the baselines    
$L_m$ and $L_c$  are correlated:  they are  
determined by the zenith angle, $\Theta_{\nu}$ of neutrino trajectory:  
\be
L_m = R |\cos \Theta_{\nu}| - L_c/2, ~~~~ L_c = 2 \sqrt{R^2 \cos^2 \Theta_{\nu} - (R^2 -  
R_c^2)}.  
\label{baselines}
\ee
Here  $R = 6370$ km is the radius of the Earth and $R_c = 3486$ km  
is the radius of the core.  Using eqs.  
(\ref{phases2}), (\ref{baselines}) we obtain that the condition 
 (\ref{magic-gen3}) is satisfied for  
\be
|\cos \Theta_{\nu}|_{magic} \approx 0.88,~~~ {\rm and} ~~~ |\cos \Theta_{\nu}|_{magic} \approx 
0.98. 
\label{zen-angle}
\ee
(For this estimation we took the potentials according to the average 
densities in the mantle and in the core $\rho_m = 4.5$ g/cm$^3$ and $\rho_c = 11.5$ 
g/cm$^3$.) Thus, there are 
two magic trajectories for neutrinos that cross the core.  
They correspond to the baselines $L = 2 R |\cos \Theta_{\nu}|$, 
$L_{magic}  = 11 210$ km and $L_{magic}  = 12 485$ km.  
For these baselines the  total oscillations phases,    
$\phi_c + 2\phi_m$, equal  $ \sim 4\pi$  and $\sim 6\pi$.  
 
These new magic baseline could of interest for the atmospheric neutrino studies.  


\section{More magic baselines}

The interference, and consequently, dependence on the CP-phase 
vanish also when $A_A = 0$.  
In this case the probability is determined by 
the solar amplitude: 
\be
P(\nu_e\to \nu_e) \approx |\cos \theta_{23} A_{S}|^2. 
\ee
Now the probability is determined by the solar parameters and can be used    
for measurements of $\theta_{12}$ as well as 
and $\theta_{23}$.

In the constant density approximation, up to the phase 
factor,  the atmospheric amplitude equals  
\be 
A_A = \sin 2 \theta_{13}^m \sin \frac{\phi_A^m}{2}.  
\label{ampl-A}
\ee
(The phase factor $\exp(- i\Delta_{13}L/2)$  
is omitted in  (\ref{ampl-A}). This factor should be restored 
if both solar and atmospheric amplitudes are non-zero.)  
So, the condition $A_A = 0$ gives the 
``magic'' baseline (integer of the oscillation length  in matter) 
\be
L_{magic} = 
n l_\nu [(\cos 2 \theta_{13}  - l_\nu/l_0)^2 + \sin^2 2 \theta_{13}]^{-1/2}, 
\label{length13}
\ee
$n= 1, 2,...$ and here  $l_\nu = 4\pi E/\Delta m_{13}^2$. 

An interesting range of energies is below 3-4 GeV where 
the amplitude $A_{S}$ is not suppressed too strongly by matter effects. 
Here, however, the ``magic'' baseline 
strongly depends on neutrino energy. Therefore, a narrow energy 
neutrino beam should be employed or reconstruction of the neutrino energy 
should be done  to suppress the interference and dependence of the probability 
on $\delta$.  For $E_\nu = 1$ GeV the magic baselines  are 
1080 km  (n = 1), 2070 km (n = 2), 3250 km (n = 3),  {\it etc.}.  
For $L = 3700$ km that maximizes the 
solar amplitude, the ``magic'' energies are at 1.1 GeV, 1.5 GeV and 
2.5 GeV.  \\


\section{Conclusion} 

The magic baseline (in the first approximation) 
is an integer of the refraction length.  
At high energies the latter approximately equals the oscillation length in matter. 
Therefore at the ``magic'' baseline the phase of oscillations 
driven by the solar mass splitting is $2\pi$, and consequently the 
solar amplitude vanishes in the transition probability. 
The interference of amplitudes, and consequently, dependence of probability 
on the CP-phase disappear. Defining the magic baseline as the distance  
that corresponds to vanishing solar amplitude, 
we have estimated various corrections to the equality $L_{magic} = l_0$. 
The magic lengths in the non-uniform medium were discussed,  and 
it is found that two additional magic trajectories exist for 
neutrinos crossing the core. 
We also discussed features of the  magic baselines  that  correspond to zeros of the  
atmospheric amplitude.

\section*{Acknowledgments}

I would like to thank H. Minakata who rose the question 
about physical meaning of the ``magic'' baseline, T. Schwetz for discussion  and 
E. Kh. Akhmedov for  encouragement to publish this  material. 
I am grateful to M. Maltoni whose comments forced me to expand 
discussion in the second version of the paper.



\begin{thebibliography}{99}



\bibitem{barger}
    V.~Barger, D.~Marfatia and K.~Whisnant,
    Phys.\ Rev.\ D {\bf 65} (2002) 073023
    [arXiv:hep-ph/0112119].



\bibitem{winter}
    P.~Huber and W.~Winter,
    Phys.\ Rev.\ D {\bf 68} (2003) 037301
    [arXiv:hep-ph/0301257].

\bibitem{more} see {\it e.g.},
    P.~Huber, M.~Lindner, M.~Rolinec and W.~Winter,
    Phys.\ Rev.\ D {\bf 74} (2006) 073003
    [arXiv:hep-ph/0606119];
    
    A.~Blondel, A.~Cervera-Villanueva, A.~Donini, P.~Huber, M.~Mezzetto and P.~Strolin,
    Acta Phys.\ Polon.\ B {\bf 37} (2006) 2077
    [arXiv:hep-ph/0606111];


    A.~Donini, E.~Fernandez-Martinez, D.~Meloni and S.~Rigolin,
    Nucl.\ Phys.\ B {\bf 743} (2006) 41
    [arXiv:hep-ph/0512038];

    P.~Huber, M.~Lindner, M.~Rolinec and W.~Winter,
    Phys.\ Rev.\ D {\bf 73} (2006) 053002
    [arXiv:hep-ph/0506237];

    R.~Gandhi, P.~Ghoshal, S.~Goswami, P.~Mehta and S.~Uma Sankar,
    arXiv:hep-ph/0506145;

    R.~Gandhi, P.~Ghoshal, S.~Goswami, P.~Mehta and S.~Uma Sankar,
    Phys.\ Rev.\ D {\bf 73} (2006) 053001
    [arXiv:hep-ph/0411252];

    E.~K.~Akhmedov, R.~Johansson, M.~Lindner, T.~Ohlsson and T.~Schwetz,
    JHEP {\bf 0404} (2004) 078
    [arXiv:hep-ph/0402175];

    W.~Winter,
    Phys.\ Lett.\ B {\bf 613} (2005) 67
    [arXiv:hep-ph/0411309].


\bibitem{huber}
    P.~Huber,
    J.\ Phys.\ G {\bf 29} (2003) 1853
    [arXiv:hep-ph/0210140].

    
\bibitem{w}
L.~Wolfenstein,
    Phys.\ Rev.\ D {\bf 17} (1978) 2369.


\bibitem{MSW} S. P. Mikheyev and A. Yu. Smirnov, 
Yad. Fiz. {\bf 42} (1985) 1441, [Sov. Jour. Nucl. Phys. {\bf 42}  (1985) 913]. 



\bibitem{ams}
  E.~K.~Akhmedov, M.~Maltoni and A.~Y.~Smirnov,
  Phys.\ Rev.\ Lett.\  {\bf 95} (2005) 211801
  [arXiv:hep-ph/0506064].

\end{thebibliography}
\end{document}